\documentclass[aps,twocolumn,prd,showpacs,nofootinbib]{revtex4}
\usepackage{amsmath}
\usepackage{graphicx}
\usepackage{dcolumn}
\usepackage{bm}
\usepackage{amssymb}
\usepackage{latexsym}

\def\be{\begin{equation}}
\def\ee{\end{equation}}
\def\ba{\begin{eqnarray}}
\def\ea{\end{eqnarray}}

\begin{document}

\title{Oscillating Quintom and the Recurrent Universe }

\author{Bo Feng$^1$, Mingzhe Li$^1$, Yun-Song Piao$^{1,2,3}$, and Xinmin Zhang$^{1}$ }
\affiliation{$^{1}$Institute of High Energy Physics, Chinese
Academy of Sciences, P.O. Box 918-4, Beijing 100039, P. R. China}
\affiliation{$^{2}$Interdisciplinary Center of Theoretical
Studies, Chinese Academy of Sciences, P.O. Box 2735, Beijing
100080, China} \affiliation{$^{3}$College of Physical Sciences,
Graduate School of Chinese Academy of Sciences, YuQuan Road 19{\rm
A}, Beijing 100049, China}

\begin{abstract}
Current observations seem to mildly favor an evolving dark energy
with the equation of state getting across -1. This form of dark
energy, dubbed Quintom, is studied phenomenologically in this
paper with an oscillating equation of state. We find oscillating
Quintom can unify the early inflation and current acceleration of
the universe, leading to the oscillations of the Hubble constant and a
recurring universe. Our oscillating Quintom would not lead to a
big crunch nor big rip. The scale factor keeps increasing from one
period to another and leads naturally to a highly flat universe.
The universe in this model recurs itself and we are only staying
among one of the epochs, in which sense the coincidence problem is
reconciled.

\end{abstract}


 \maketitle

In 1998, two groups \cite{Riess98,Perl99} independently showed the
accelerating expansion of our universe, which established the
existence of dark energy where the equation of state is less than
$-1/3$ today. The simplest form of dark energy is the cosmological
constant. A cosmological constant which leads to current
acceleration would encounter many theoretical problems, such as the
fine-tuning problem and the coincidence problem\cite{weinberg}. A
light scalar field of quintessence\cite{pquint,quint,Wangtrk}
which evolves with time is to some extent likely to resolve the
coincidence problem. The model of
phantom\cite{phantom} has also been put forward which leads to an
equation of state $w\leq -1$. Normally quintessence with a canonic
kinetic term can only have $w\ge -1$, meanwhile
k-essence\cite{k-essence} can have both $w\ge -1$ and $w<-1$. For
the cosmological constant and many quintessence models the
event horizon would lead to a potential incompatibility with the string
theory, meanwhile for models where $w<-1$ one would get
the Big Rip\cite{CKW} of the universe.

The accumulation of the current observational data has opened a
robust window for probing the recent behavior of dark energy.
Measurements from type Ia Supernova (SNe Ia), the Cosmic Microwave
Background (CMB) Radiation, Large Scale Structure (LSS), weak
lensing, clusters  and galaxies all contain the imprints of dark
energy. Specifically the recently released first year Wilkinson
Microwave Anisotropy Probe (WMAP) measurement\cite{Spergel03}, the
Sloan Digital Sky Survey (SDSS) measurement of the three-dimensional
power spectrum\cite{0310723} and most importantly, the recent
discovery of 16 SNe Ia \cite{Riess04} with the Hubble Space
Telescope during the GOODS ACS Treasury survey, together with former
SNe Ia data have provided the most precise up-to-date measurements
of dark energy. Many other sources have also been studied to make
crosschecks and reveal new physics on the dark energy and the
concordance cosmological model. Recently many authors in the
literature have fitted the behavior of dark energy using various
parameterizations. The best fit value of the equation of state is
found to be less than $-1$ when using SNe Ia
\cite{Riess04,sahni,cooray,FWZ,CP0304,DR04} or the X-ray mass
fraction data\cite{Allen04} with a cosmological constant still well
within the central regions\footnote{We get the similar results when
using the recent data of radio galaxies\cite{dd03} as well as the
gold samples of SNe Ia data\cite{Riess04}}. It has also been pointed
out by many authors that an evolving dark energy is indeed more
favored than that with a constant equation of state. With the
simplest linear parametrization of the equation of state $w$:
\begin{equation}
w(z)=w_0+w'z ,
\end{equation}
 SNe Ia alone favors
a $w$ larger than $-1$ in the recent past and less than $-1$ today,
disregard using the prior of a flat universe
\cite{Riess04,sahni,cooray,FWZ,CP0304} or not\cite{DR04}. When other
"longer-armed" observations, e.g. CMB and LSS data, are also taken
into account, one often takes the risk of poor parametrizations.
Taking the linear parametrization as a concrete example, it gives a
good approximation for the late behavior of the dark energy.
However, at large redshift the amplitude of $w$ increases too much
and hence CMB and LSS data would restrict $w'$ to be near
zero\cite{Riess04,wangy04}, spoiling the recent behavior of dark
energy. Such a correlation also happens for other parametrizations.
The most proper way of measuring the recent behavior of the dark
energy is the principal component analysis\cite{Huterer}. Huterer
and Cooray\cite{cooray} based on this method and developed a simple
uncorrelated and nearly model-independent band power estimates of
$w$ and its density as a function of the redshift.  By fitting to
the recent SNe Ia data they found a marginal (2-$\sigma$) evidence
for $w(z) < -1$ at $z < 0.2$. Even though SNe Ia data are still
currently limited, such a method is potentially very powerful. The
parametrization of the dark energy remains a paradox today,
especially when confronted with observations \cite{Linde04}.
However, in any case if a time-varying dark energy model
significantly reduces the $\chi^2$ value and is statistically
significant, it gives strong implications for dark energy
metamorphosis. In Refs.\cite{Corasaniti,Hannestad} the authors use
their physical parametrizations of the dark energy and fit the
models to the fully CMB, LSS and SNe Ia data set, they also find
that an evolving dark energy is favored indeed with $w(z) < -1$
today and $w(z) > -1$ in the recent past. However, one would have to
introduce several additional parameters to give a physical
interpretation for the behavior of the dark energy, resulting into a
very low statistical significance of an evolving dark energy. Most
of the works in the literature have not taken into account the
possibly non-negligible contribution of weak lensing\cite{wangyls}
to SNe Ia. When the flux averaging method \cite{wangyflux,wangy03}
has been applied the preference of a nonzero $w'$ is much reduced.
In Ref.\cite{FWZ} we have also applied the linear parametrization to
the X-ray mass fraction data only and find that a negative $w'$ is
favored.

If the requirement of an evolving dark energy with $w$ getting
across $-1$ still holds on with the accumulation of observational
data, this would be a great challenge to the current cosmology. In the
quintessence model, the equation of state $w=p/\rho$ is always in
the range $-1\leq w\leq 1$ for $V(Q)>0$. Meanwhile for the phantom
which has the opposite sign of the kinetic term compared with the
quintessence in the Lagrangian, one always has $w\leq -1$. Neither
the quintessence nor the phantom alone can fulfill the transition
from $w>-1$ to $w<-1$ and vice versa. Although for
k-essence\cite{k-essence} one can have both $w\ge -1$ and $w<-1$,
it has been lately considered by Ref\cite{Vikman} that it is very
difficult for k-essence to get $w$ across $-1$ during evolving.
In Ref.\cite{FWZ} we proposed a new scenario of
dark energy dubbed Quintom. Quintom describes those forms of dark
energy which can get $w$ cross $-1$ during the time evolution.

When dark energy is not merely the cosmological constant it
gives rise to many interesting consequences. There are possibly
some connections between the early inflation and the current
acceleration of our universe. Peebles and Vilenkin have put
forward a model of quintessential inflation to unify the two
epochs
\cite{PVQinf}. Dodelson, Kaplinghat and
Stewart gave a simple form of oscillating quintessence which can
provide a natural solution to the coincidence
problem\cite{DSPRL}. In this paper we propose a
phenomenological
model of oscillating Quintom and study its implications in
cosmology. We will show that in our model two
accelerating epochs are related and the coincidence
problem can be alleviated.

The phenomenological Quintom model we consider in this paper takes
the following equation of state,
\begin{equation}\label{osceq}
w(\ln a)=w_0 + w_1 \cos[A \ln (a/a_c)],
\end{equation}
where $a$ is the scale factor, $w_0$, $w_1$, $A$ and $a_c$ are
parameters which we will take specific values for the detailed
study below. If for a sufficiently large period the equation of
state remains unchanged during the late time evolving it would fit
all current data well for $w\lesssim -1$. Defining $X(\ln
a)\equiv\rho_X(\ln a)/\rho_X(0)$ where the subscript $X$ stands
for Quintom and $'0'$ for today, one gets:
\begin{equation}\label{oscrho}
X(\ln a)=a^{-3(1+w_0)}\exp\{{\frac{-3 w_1}{A}[\sin (A \ln
\frac{a}{a_c}) +\sin (A \ln a_c) ] \}}~.
\end{equation}

For a numerical discussion we fix the model parameters and consider
a specific $w$ given below
\begin{equation}\label{osceq1}
w(\ln a)=-1 - 1.5 \cos(0.032 \ln a - \frac{4 \pi}{9}).
\end{equation}
In this model the $w$ at present time is $w=-1.26$ ( we have checked
that if tuning $a_c$ to get $w$ more closer to $-1$ the picture on
the evolution of the universe will not be changed from our current
choice of the parameters), which is well within the limit of current
observations. In Fig.1 we show the evolution of the universe filled
with only the matter of the oscillating Quintom. As we've fixed
$w_0=-1$, one can see from Eq.\ref{oscrho} that the energy density,
hence the Hubble parameter $H$ simply oscillates with a long period.
Since the time is a simple integration of $d \ln a/H$ and $X(\ln a)$
evolves periodically with $\ln a$, time increases equally for every
span of $\ln a$. One can also easily find from Fig.1 that as $X$ is
an integration of $w(\ln a)$ they differ in the phase of
oscillation. Since the Hubble parameter increases gradually during
phantom-dominated phase of dark energy, this will make the universe
evolve into a regime with high energy again \footnote{This owes to
the special features of phantom, in which the null energy condition
is violated. For the oscillation of state equation without $w <-1$,
this case does not occur. }, which may be regarded as the recurrence
of an early phantom inflation \cite{PZhang} of our universe, (see
\cite{PZhou} for a earlier study) where the initial perturbation in
the horizon will exit the horizon, and reenter the horizon after the
transition to a late-time expanding phase with the ordinary matter
contents. Thus our model unifies the inflation and dark energy into
a single process, in which the universe evolves periodically. While
we live in one period, the present acceleration with $w <-1$ is just
a start of the phantom inflation for the next period of universe.
During this process, all ordinary forms of matter and radiation are
diluted by inflation. But in the next period new structures will be
formed again by the fluctuations generated during the present
acceleration, {\it i.e.} phantom inflation \cite{PZhang} (see
\cite{PZhang2, Piao} for a detailed investigation of primordial
perturbations from phantom phase). The universe can also be reheated
with help of the reheating mechanisms such as instant
preheating\cite{preheating} and curvaton reheating\cite{FL02}.

\begin{figure}[htbp]
\begin{center}
\includegraphics[scale=0.5]{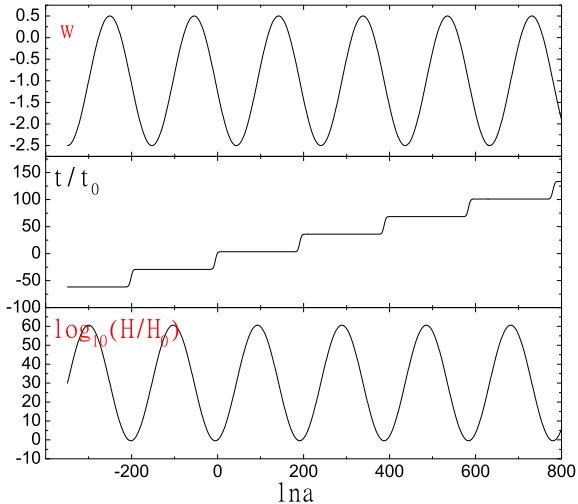}
\caption{Cosmological consequences with only an oscillating
Quintom $w(\ln a)=-1 - 1.5 \cos(0.032 \ln a - \frac{4 \pi}{9})$.
\label{fig:fig1}}
\end{center}
\end{figure}

We show now that our model works the same as described above when
including matter and radiation. Assuming $\Omega_m=0.3$ and
$\Omega_X=0.7$ today with the Hubble parameter $h=0.7$ ( this
determines the abundance of radiation ) and a negligible curvature,
we show in Fig.2 the evolution of the universe for 1-$period$.
 Here we assume that the radiation
starts to dominate our universe promptly at the moment when $\ln
(a/a_0) = -60$ ( $a_0$ is set to be unity today ). From Fig.2 one
can see that firstly the universe is just like today with
$\Omega_X=0.7$ and $w=-1.26$, with time $w$ evolves downward,
$\rho_X$ increases and $\rho_m$ decreases, then the universe begins
inflating. When $\ln a \sim -150$, $w$ starts to increase. And at
$\ln a \sim -90$ we have $w\sim -\frac{1}{3}$, inflation stops and
the energy density continues dropping. At $\ln a = -60$ the universe
enters radiation domination epoch. $\rho_X$ starts to increase
recently and begins to dominate the universe with $\Omega_X=0.7$
today. The universe starts the inflation since today and is
recurring into the next period.

\begin{figure}[htbp]
\begin{center}
\includegraphics[scale=0.5]{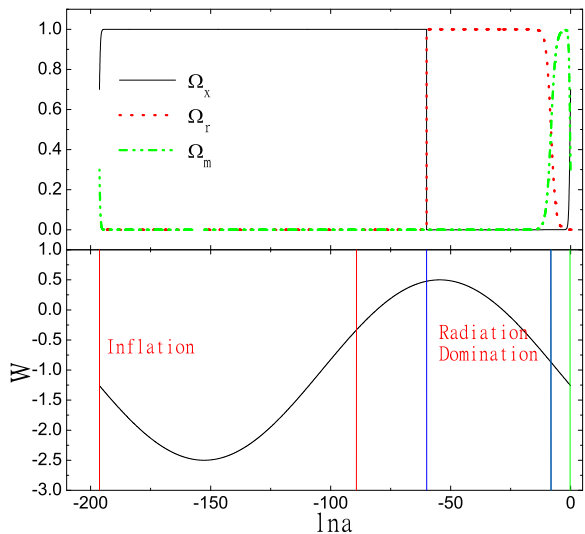}
\caption{Cosmological consequences with an oscillating Quintom
$w(\ln a)=-1 - 1.5 \cos(0.032 \ln a - \frac{4 \pi}{9})$ and
$\Omega_m=0.3$, $\Omega_X=0.7$ and $h=0.7$ today, within only one
period. \label{fig:fig2}}
\end{center}
\end{figure}

In Fig. 3 we show the evolution of the universe with multi
periods. For each period we assume that the radiation starts
domination around the same time. As radiation and Quintom
domination epoch contribute little to time compared with when
matter is non-negligible, one can see from Fig. 3 that time only
increases equally around every short epoch of matter domination.

\begin{figure}[htbp]
\begin{center}
\includegraphics[scale=0.5]{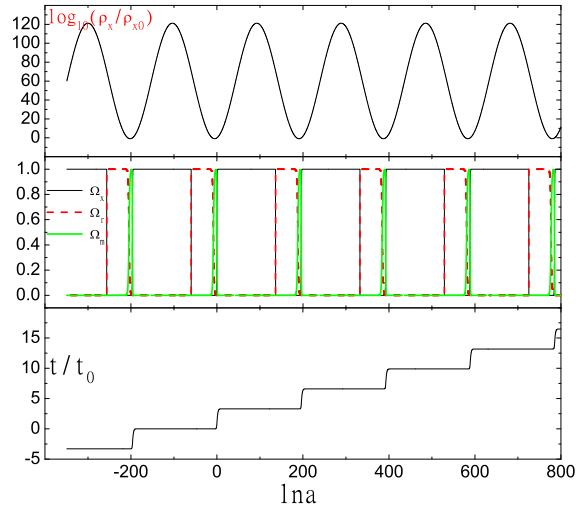}
\caption{Cosmological consequences with an oscillating Quintom
$w(\ln a)=-1 - 1.5 \cos(0.032 \ln a - \frac{4 \pi}{9})$ and
$\Omega_m=0.3$, $\Omega_X=0.7$ and $h=0.7$ today, within several
periods. \label{fig:fig3}}
\end{center}
\end{figure}

We should point out that for our phenomenological parametrization in
Eq.(\ref{osceq}) and Eq.(\ref{osceq1}) to realize our recurrent
picture of the universe, the value of $w_0$ has to be extremely
close to -1, otherwise the next and previous "cycles" would be
different from the current one and we would also be encountered with
too high or low inflation energy scales, as can be clearly seen in
Eq.(\ref{oscrho}). While for the other several parameters less fine
tunings are needed: the period is determined by $A$ and the
variation of the energy density is by $w_1$ and $A$, and finally,
the value of $a_c$ somewhat determines the present value of $w$. To
realize the recurrence picture one requires that inflationary epoch
is satisfactory and hence we have $2\pi/A \sim 100$, on the other
hand we need $H_{inf}/M_{Pl}<10^{-5}$ to satisfy the primordial
gravitational wave constraints, these will put considerable
constraints on $w_1$. For our dynamical model of dark energy as the
period of oscillation is required to be quite large, the crossing of
the cosmological boundary appears at some high redshifts where dark
energy is typically negligible compared with the background matter
components, where such a crossing behavior is hard to be detected by
the observations. Nevertheless such a crossing is crucial to realize
our picture of oscillating quintom. The dynamics of dark energy is
not well imprinted on the redshift range covered by current SNe Ia
surveys, where a running of the equation of state $d w /dz $ would
be typically of order 0.01, this is different from the model of
Quintom investigated earlier where a significant dynamical dark
energy behavior displays in a very low and short redshift
range\cite{FWZ}. For the specified example in Eq.(\ref{osceq1}) the
running of $w$ is $\sim 0.02$, which is difficult to be excluded by
future observations like SNAP\cite{Aldering:2004ak,SNAPweb}, Joint
Efficient Dark-energy Investigation
(JEDI)\cite{Crotts:2005eu,JEDIweb} or Dark Energy Survey
(DES)\cite{Abbott:2005bi}. On the other hand, if a significant
dynamical behavior of dark energy were verified by future
observations, this would be a strong argument to rule out our model.
Again we should point out that Eq.(\ref{osceq1}) is one of the
typical examples where the current equation of state is quite
adjustable which does not affect our picture. To satisfy the current
observations we need the equation of state close to $-1$ today,
which in turn requires in Eq.(\ref{osceq}) the term $w_1 \cos[A \ln
a_c]$ to be close to zero.

During the phantom-dominated phase, the energy density increases as
well as the Hubble parameter. If there is not a transition, the exit
to observable universe is hardly possible, which is also a problem
puzzling phantom inflation \footnote{The exit may be implemented by
using growing wormholes and ringholes \cite{GD}, which can be
regarded as a connection between the phantom phase and late-time
expanding phase, see \cite{GD2} for a further discussion.}. But in
our model it is avoided by the oscillation of $w$. In some sense,
our model is similar to the Pre Big Bang scenario \cite{GV}( see
\cite{V} for a review) and the recently proposed cyclic model
\cite{KOS} where the primordial perturbations are seeded by the
contraction phase with the increase of the energy density. To exit
and enter late-time observable universe, a non singular bounce has
to be required. But different from those which lead to the big
crunch of the universe, in our model the universe is always in the
expanding phase with periodicity. In this case the ``bounce"
actually means a transition from one expanding phase to another.

In summary we have proposed in this paper a phenomenological model of
Quintom with an oscillating equation of state. Such a model
naturally unifies the early inflation and the current acceleration of the
universe, leading to oscillations of the Hubble constant and a
recurring universe. The model of oscillating Quintom does not lead
to a big crunch nor big rip. The universe just recurs itself with
the scale factor increasing always and we are only staying among
one of the epochs, in which sense the coincidence problem is
somewhat reconciled and $\Omega_k$ is predicted to be highly close
to zero\cite{relev}. Further studies on model building, perturbation and
reheating are currently under investigation\cite{prepare}.

{\bf Acknowledgements:} We thank the anonymous referee for helpful
comments and suggestions. This work is supported in part by
National Natural Science Foundation of China under Grant Nos.
10405029, 90303004 and 19925523 and by Ministry of Science and
Technology of China under Grant No. NKBRSF G19990754.

\end{document}